\documentclass[aps,prb,showpacs,showkeys,amsmath,amssymb,byrevtex]{revtex4}
\usepackage{graphics}
\DeclareMathOperator{\dlim}{\mathrm{d-lim}}
\DeclareMathOperator{\tr}{\mathrm{Tr}}
\begin{document}
\title{On~the~\mbox{Kondo}~problem and~thermodynamics of~dilute magnetic alloys}
\author{Jan \surname{Ma\'{c}kowiak}}
\email[]{ferm92@fizyka.umk.pl}
\affiliation{Instytut Fizyki, Uniwersytet M.~Kopernika, ul.~Grudzi\c adzka~5/7, 
87--100~Toru\'n, Poland}
\date{\today}
\begin{abstract}
  An~argument is~given showing that~\mbox{Coulomb} attraction between conduction
  electrons and~impurity ions in~a~dilute magnetic alloy~(DMA) can~be~disregarded,
  provided the~system's inverse temperature~$\beta$ is~replaced by~an~effective
  inverse temperature~$t < \beta$. This~replacement allows to~remove the~singularity
  in~\mbox{Kondo's}~expression for~DMA impurity resistivity and~extend his~theory
  to~$0 \text{ K}$. The~extended \mbox{Kondo}~formula agrees
  with~experimental data on~resistivity of~CuFe in~the~range of~low temperatures
  and~in~the~neighbourhood of~the~resistivity minimum. 
  
  Using an~asymptotic solution of~the~thermodynamics of~a~dilute \mbox{s-d}~system 
  at~inverse temperature~$t$, the~impurity thermodynamic functions are~derived 
  and~shown to~provide good agreement with~experimental data on~CuFe, CuCr
  and~$\text{(LaCe)Al}_2$ alloys in~the~low-temperature range. The~magnitude
  of~these~functions agrees with~experiment and~does~not~require rescaling
  as~in~previous \mbox{s-d}~theories. Nonlinear dependence of~CuFe heat~capacity
  on~impurity concentration has~been~accounted~for the~first time.
\end{abstract}
\maketitle
\section{\label{sec:1}Introduction}
The~anomalous thermal behaviour of~dilute magnetic alloys~(DMA) has~been the~subject
of~extensive experimental and~theoretical research over the~past decades. The~main
stream of~theoretical 
investigations~(e.g.~Refs.~\onlinecite{kondo,bloomfield,hepp,wilson,rajan,andrei,%
filyov2,wiegmann,hewson,yosida,rudavskii}) has~focused on~the~construction 
of~a~conductivity~theory and~thermodynamics of~DMA on~the~grounds
of~the~\mbox{s-d}~Hamiltonian~$H_{\text{s-d}}$ introduced
by~\mbox{Kasuya}\cite{kasuya}.

\mbox{Kondo's}~theory of~DMA impurity resistivity~$\Delta \rho$\cite{kondo}
has~successfully explained the~experimentally observed dependence of~$\Delta \rho$
on~impurity concentration~$c$ and~temperature~$T$ in~the~vicinity of~the~resistivity
minimum. Unfortunately, the~theory fails at~$T=0$, where~\mbox{Kondo's}~expression
for~$\Delta \rho$ exhibits a~logarithmic singularity. The~question of~removing
this~singularity is~known as~the~``\mbox{Kondo}~problem''.

Substantial progress in~understanding the~anomalous~DMA~thermodynamics was~made
by~\mbox{Andrei}~et~al.\cite{rajan,andrei} who~solved~$H_{\text{s-d}}$ thermodynamics
for~an~\mbox{s-d}~system with indistinguishable impurities and~point 
\mbox{s-d}~interaction. Their~rescaled thermodynamic functions,
agree, up~to~a~small error, with~experimental impurity specific heat
and~magnetization of~$\text{(LaCe)Al}_2$. The~solution
found~in~Refs.~\onlinecite{rajan,andrei} yields universal single-impurity curves
for~each thermodynamic function corresponding to~a~given value of~impurity spin,
which~are~independent of~impurity concentration~$c$. 

The~shape of~experimental plots of~DMA~thermodynamic functions, in~general, varies
slowly with~$c$~(e.g.~Refs.~\onlinecite{kondo,franck,felsch}), meaning 
that~their~dependence on~$c$ is~nonlinear. This~type of~dependence 
has~been~accounted~for by~theory only in~exceptional 
cases~(e.g.~Refs.~\onlinecite{kondo,souletie}).

A~different solution of~$H_{\text{s-d}}$ thermodynamics, which~treats a~dilute
\mbox{s-d}~system (with a~smeared \mbox{s-d}~interaction) containing arbitrarily 
positioned distinguishable impurities and~yields nonlinear dependence of~thermodynamic 
functions on~$c$, was~presented by~the~author in~Refs.~\onlinecite{m2,m3}. 
The~first quantization Hamiltonian~$H_{\text{s-d}}$ in~this~approach is
\begin{equation}
        H_{\text{s-d}}^{(n,M)}=A^{(n)}\left( H_0^{(n)} + g^2 \sum_{\alpha=1}^M 
	\sum_{i=1}^n U(\mathbf{R}_{\alpha}-\mathbf{r}_i)\otimes S_{z\alpha}\sigma_{zi} 
	\right),     
\label{H_sd}
\end{equation}
$n$ $(M)$ denoting the~number of~electrons (impurities), $A^{(n)}$~the~antisymmetrizer 
with~respect~to electron variables with~indices~$i=1,\ldots , n $, and
\begin{equation}
        H_0^{(n)}=-\sum_{i=1}^n \frac{\hslash^2}{2m} \mathit{\Delta}_i.
\label{H0}
\end{equation}
$\mathbf{R}_{\alpha}$ denotes the~position vector of~the~$\alpha$th~impurity, 
$S_{z\alpha}$~its spin operator and~$\mathbf{r}_i$, $\sigma_{zi}$~stand for~the
respective quantities of~the~$i$th~electron. $U \ge 0$~represents any~sufficiently 
regular function depending on~$|\mathbf{R}_{\alpha}-\mathbf{r}_i|$, which~allows 
application of~the~\mbox{Feynman}-\mbox{Kac}~theorem to~$\exp [-\beta
H_{\text{s-d}}^{(n,M)}]$\cite{simon}. This~theorem was~applied
in~Ref.~\onlinecite{m2} to~derive an~upper and~lower bound to~the~system's
free~energy per~electron,
\begin{equation*}
  f(H_{\text{s-d}}^{(n,M)},\beta) := - (n\beta)^{-1} \ln \tr \exp [-\beta 
  H_{\text{s-d}}^{(n,M)}],
\end{equation*}
and~to~prove that~the~two bounds coaleesce in~the~limit of~small~$c \to 0$~($\dlim$)
and, as~a~consequence, that
\begin{equation}
  \lim_{n \to \infty} \dlim f(H_{\text{s-d}}^{(n,M)},\beta) = \lim_{n \to \infty} 
  \dlim \min_{\xi,\eta} f(h^{(n,M)}(\xi,\eta),\beta) = \lim_{n \to \infty}
  f(H_0^{(n)} A^{(n)},\beta),
  \label{free_en}
\end{equation}
where~$h^{(n,M)}(\xi,\eta)$ is~the~mean-field~Hamiltonian
of~an~\mbox{s-d}~system~$S_0$ with~separated electron and~impurity variables. Both
the~electron subsystem~$S_{\text{e}}$ and~the~impurity subsystem~$S_{\text{imp}}$
of~$S_0$ consist of~noninteracting particles. According~to~Eq.~\eqref{free_en},
$h^{(n,M)}(\xi,\eta)$~is~almost thermodynamically equivalent to~$H_{\text{s-d}}$
in~the~extreme dilute limit.

The~1-electron Hamiltonian of~$S_{\text{e}}$ has~the~form
\begin{equation}
        h_{\text{e}}^{(1,M)}(\xi,\eta)=\tilde{h}_{\text{e}}^{(1,M)}(\xi,\eta)+
	\frac{1}{2} M (\xi^2-\eta^2)\mathbb{I}
        \label{h_e}
\end{equation}
with
\begin{equation}
        \tilde{h}_{\text{e}}^{(1,M)}(\xi,\eta)=H_0^{(1)} - g\sqrt{n} (\xi-\eta) 
	\sum_{\alpha=1}^M U_{\alpha}^{(1)}\otimes \sigma_z^{(1)}.
        \label{h_e_tilde}
\end{equation}
$U_{\alpha}^{(1)}$~denoting the~multiplication operator by~$U(\mathbf{R}_{\alpha}
-\mathbf{r}_i)$ and~$\eta(\xi)=\xi - f_2(\xi)$, where
\begin{equation}
        f_2(\xi)= - \frac{g}{\sqrt{n}}\left\langle S_z \right\rangle_{h_{
	\text{imp}}^{(1)}}\text{,} \qquad 
	\left\langle B \right\rangle_{h} :=  \frac{\tr(B\exp (-\beta h))}{\tr
	\exp (-\beta h)},
        \label{f2}
\end{equation}
whereas~$h_{\text{imp}}^{(1)}$ is~the~1-impurity Hamiltonian of~$S_{\text{imp}}$:
\begin{equation}
        h_{\text{imp}}^{(1)}(\xi)=g\sqrt{n}\xi S_{z\alpha} + 
	\frac{1}{2} g^2 S_{z\alpha}^2.
        \label{h_imp}
\end{equation}
The~necessary condition for~the~minimum in~Eq.~\eqref{free_en} takes the~form
\begin{equation}
  \xi=f_3(\xi)
  \label{xi}
\end{equation}
with
\begin{equation}
        f_3(\xi) := f_1\left(f_2(\xi)\right)+f_2(\xi),
\end{equation}
\begin{equation}
        f_1(\xi) := g\sqrt{n}\left\langle \Gamma_1^n U_{\alpha}^{(1)}\sigma_z^{(1)}
	\right\rangle_{n\Gamma_1^n \tilde{h}_{\text{e}}^{(1,M)}(\xi,0)},
        \label{f1}
\end{equation}
\begin{equation}
        \Gamma_1^n B^{(1)} := A^{(n)}\left( B^{(1)}\otimes \mathbb{I}^{(n-1)} \right)
	A^{(n)},
\label{gamma}
\end{equation}
$n \Gamma_1^n h^{(1)}$~denoting the~Hamiltonian of~$n$~noninteracting fermions
with~the~1-fermion Hamiltonian~$h^{(1)}$~(cf.~Ref.~\onlinecite{kummer}).

The~mean-field thermodynamics founded on~Eq.~\eqref{free_en} was~used
in~Ref.~\onlinecite{m3} to~explain the~presence of~the~impurity heat~capacity peak
of~CuCr and~$\text{(LaCe)Al}_2$ in~the~vicinity
of~the~\mbox{Kondo}~temperature~$T_K$. In~contradistinction to~earlier
papers~(e.g.~Refs.~\onlinecite{rajan,andrei,anderson1}) scaling procedures
were~not~used and nonlinear dependence of~the~CuCr peak's shape on~$c$
was~taken into~account.

One~of~the~shortcomings of~DMA~theories founded on~\mbox{s-d}~type Hamiltonians
is~the~omission of~the~\mbox{Coulomb}~attraction between impurity ions and~conduction
electrons. This~problem has~been~treated in~various ways in~the~past. \mbox{Kondo}
introduced an~additive term in~his~resistivity formula\cite{kondo} to~account
for~these~interactions. In~Refs.~\onlinecite{edwards,ambegaokar} the~equilibrium
state of~an~electron~gas interacting with~impurity ions was~studied by~averaging
the~1-electron \mbox{Green's}~function over~impurity positions. By~applying
this~method to~the~1-particle equilibrium density matrix of~a~quantum particle
in~a~field of~randomly positioned wells, representing the~screened
\mbox{Coulomb}~potential at~each~impurity site, it~was~shown in~Ref.~\onlinecite{m3}
that~in~the~low temperature regime, a~gas of~such particles behaves effectively, 
with~respect to~1-particle measurements, like
a~gas of~free particles at~an~inverse temperature~$t$ related to~the~system's real
temperature~$T$ by~the~equality
\begin{equation}
  t(\delta,T) = \delta^{-1} \tanh (\delta (k_B T)^{-1})
  \label{t}
\end{equation}
where~$\delta = \frac{1}{2}\hslash \sqrt{Mu_2 m^{-1}}$, $u_2$~denoting the~2nd~derivative
at~the~well's minimum. Accordingly, the~inverse
temperature~$\beta =(k_B T)^{-1}$ of~the~\mbox{s-d}~system under consideration
will~be subsequently replaced by~$t(\delta, T + \Delta T)$ and~$\delta$, $\Delta
T$~will~be~treated as~adjustable parameters. The~shift~$\Delta T$ is~introduced
in~order to~compensate omission of~other~DMA interactions in~$H_{\text{s-d}}$.

In~the~high temperature range, $t(\delta,T)$~approaches smoothly~$\beta$,
whereas~$\lim t(\delta,T) = \delta^{-1}$ as~$T \to 0$. Replacement of~$\beta$
by~$t$ in~\mbox{Kondo's}~resistivity formula\cite{kondo}, therefore allows
to~account~for the~\mbox{Coulomb}~interactions between impurity ions and~conduction
electrons and~to~remove the~singularity in~his~theory. In~Section~\ref{sec:2}
the~resulting expression for~DMA impurity resistivity~$\Delta \rho$ is~shown to~give
a~good fit with~experimental~$\Delta \rho$ for~CuFe with~$c=110\text{ ppm}$.
\mbox{Kondo's}~expression for~the~total DMA resistivity~$\rho$ is~also shown
to~comply with~experimental~$\rho$ under this~substitution.

The~objective of~subsequent sections is~to~study the~impurity
heat~capacity~$\Delta C$, magnetization~$\Delta M$ and~susceptibility~$\Delta \chi$
of~the~mean-field~system~$S_0$ and~to~adjust the~constants which~enter
these~thermodynamic functions to~obtain the~best possible agreement with~experiment.
For~alloys with~spin~$1/2$ impurities, such as~CuFe and~$\text{(LaCe)Al}_2$,
there~is~good agreement between theory and~experiment, as~regards dependence
on~$c$ and~$T$ in~the~low-temperature range. In~particular, nonlinear dependence
of~$\Delta C$ on~iron concentration in~CuFe has~been~accounted~for successfully
the~first time. The~magnitude of~all~thermodynamic functions agrees with~experiment. 
For~the~CuCr alloys, containing spin~$3/2$ ions, agreement
is~slightly weaker, presumably due~to~the~simplicity of~the~assumed
\mbox{s-d}~interaction in~Eq.~\eqref{H_sd}, which~permits only orbital
\mbox{s-wave}~scattering~(cf.~Ref.~\onlinecite{bloomfield}).

Computations were~carried~out using Wolfram's~Mathematica~5.2.
\section{\label{sec:2}A~solution of~the~\mbox{Kondo}~problem}
In~1964 \mbox{Kondo} derived his~well known formula\cite{kondo} for~the~impurity
resistivity of~DMA:
\begin{equation}
  \Delta \rho = c \rho_A + c \rho_M \left( 1 - 3 z g^2 \varepsilon_F^{-1} \ln T
  \right)
  \label{rho}
\end{equation}
where~$\rho_A$, $\rho_M$, $z$~are constants and~$z$ is~positive for~antiferromagnetic
\mbox{s-d}~interaction. Inclusion of~lattice resistivity~$\rho_L = a T^5$ yields
\mbox{Kondo's}~expression for~the~total DMA~resistivity~$\rho$ , which~provides a~good fit
to the~$c$, $T$~dependence of~experimental data on~resistivity
of~several~DMA~(cf.~Ref.~\onlinecite{kondo}).

The~breakdown of~formula~\eqref{rho} at~$T=0$ can~be easily amended by~noting
that \mbox{Kondo's}~theory takes into~account the~\mbox{Coulomb}~attraction between
conduction electrons and~impurity ions simply by~including an~additive constant
into~the~r.h.s. of~Eq.~\eqref{rho}. From~the~viewpoint of~Eq.~\eqref{t} it~would~be
more appropriate to~replace in~Eq.~\eqref{rho} the~true inverse temperature
of~the~alloy by~$t$. For~$\Delta T = 0$, the~resulting expression for~$\Delta \rho$
then~takes the~form
\begin{equation}
  \Delta \rho = \Delta \rho_0 + \Delta \rho_1 \ln (\tanh (\frac{\delta}{k_B T}))
  \label{Delta_rho}
\end{equation}
and~is~regular at~$T=0$.

\begin{figure}
  \includegraphics{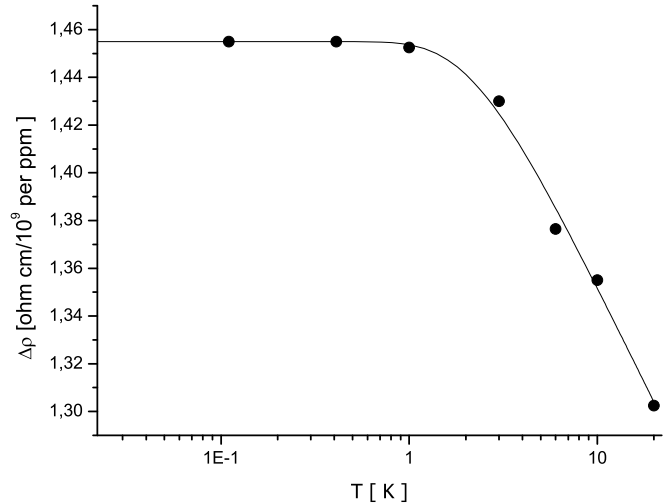}
  \caption{\label{fig:1}Impurity resistivity~$\Delta \rho$ of~CuFe,
  with~$c=22\text{ ppm}$, as~given by~Eq.~\eqref{Delta_rho}, for~$\Delta \rho_0 =
  1.455/10^9\text{ ohm per ppm}$, $\Delta \rho_1 = 0.07/10^9\text{ ohm cm per ppm}$
  and~$\delta = 2/10^4\text{ eV}$. The~points are~experimental results
  from~Ref.~\onlinecite{daybell}.}
\end{figure}

Plausibility of~formula~\eqref{Delta_rho} was~tested by~adjusting
the~constants~$\Delta \rho_0$, $\Delta \rho_1$, $\delta$~to~fit
experimental~$\Delta \rho (T)$ data for~CuFe, with~$c=22 \text{ ppm}$, plotted
in~Ref.~\onlinecite{daybell}. The~function~\eqref{Delta_rho}, for~$\Delta \rho_0 =
1.455/10^9\text{ ohm cm per ppm}$, $\Delta \rho_1 = 0.07/10^9\text{ ohm cm per ppm}$
and~$\delta = 2/10^4\text{ eV}$, is~depicted in~Fig.~\ref{fig:1}. Agreement
with~experiment is~good.

The~experimental~$\Delta \rho$ data for~CuFe with~$c=22\text{ ppm}$ are~quite
typical. Resistivity measurements of~a~variety of~DMA samples point~to the~close
similarity of~$\Delta \rho (T) / \Delta \rho (0)$ curves for~various
alloys\cite{heeger}. Formula~\eqref{Delta_rho} can~be therefore expected to~provide
good agreement with~experiment for~a~large class of~DMA.

\mbox{Hamann's}~expression for~impurity resistivity\cite{hamann}, with~$t$
replacing~$\beta$, was~also~confronted with~the~data of~Ref.~\onlinecite{daybell}.
Qualitative agreement was~found.

\begin{figure}
  \includegraphics{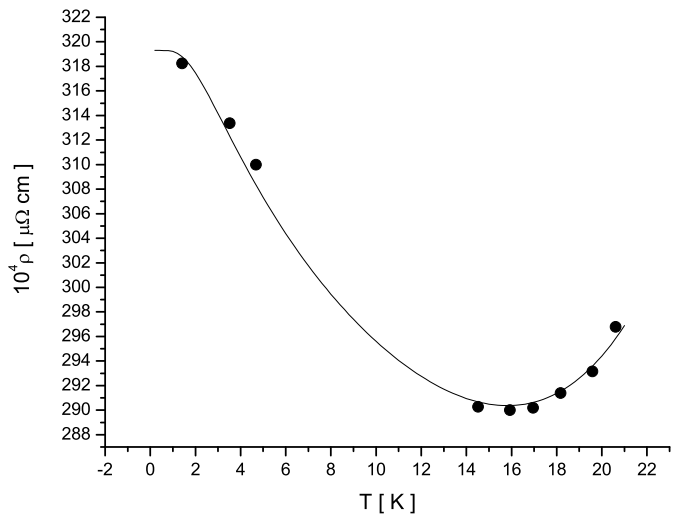}
  \caption{\label{fig:2}The~total resistivity~$\rho$ of~CuFe, with~$c=1.23/10^3$, 
  as~given by~Eq.~\eqref{rho_T_ham}, for~$\Delta \rho_0 = 319.2937/10^4\text{
  $\mu$ohm cm}$, $\Delta \rho_1 = 2/10^3\text{ $\mu$ohm cm}$, $\Delta \rho_2 = 
  1.065/10^7\text{ $\mu$ohm cm}$ and~$\delta = 2.66/10^4\text{ eV}$. The~points 
  are~experimental results from~Ref.~\onlinecite{berg}.}
\end{figure}

Performing the~substitution~$\beta\to t$ in~\mbox{Kondo's}~expression for~total DMA
resistivity~$\rho$, one obtains
\begin{equation}
  \rho(T) = \Delta \rho(T) + \Delta \rho_2 \left( \tanh(\frac{\delta}{k_B
  T})\right)^{-5}.
  \label{rho_T_ham}
\end{equation}
For~$\Delta \rho_0 = 319.2937\times 10^{-4} \mu\text{ohm cm }$, $\Delta \rho_1 = 2
\times 10^{-3} \mu\text{ohm cm }$, $\Delta \rho_2 = 1.065\times 10^{-7} 
\mu\text{ohm cm }$ and~$\delta = 2.66 \times 10^{-4}\text{ eV}$,
the~function~$\rho(T)$ provides a~good fit to~the~experimental plot
of~$\rho(T)$ for~CuFe with~$c=1.23\times 10^{-3}$~(Ref.~\onlinecite{berg})
and~is~depicted in~Fig.~\ref{fig:2}.

\begin{figure}
  \includegraphics{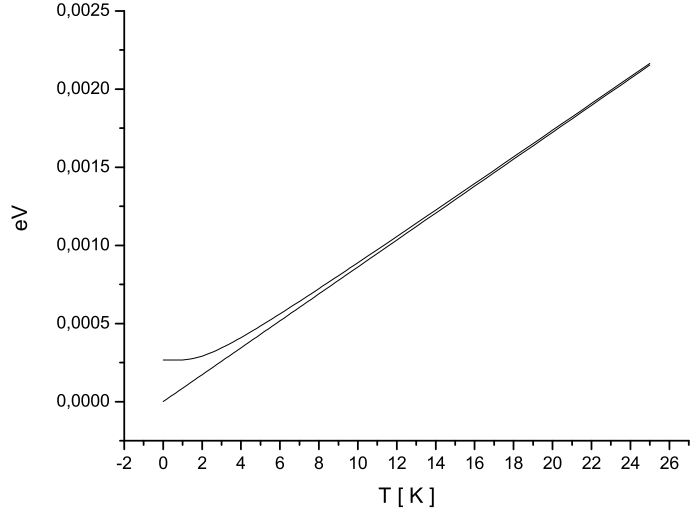}
\caption{\label{fig:3}The~graphs of~$k_B T$ and~$t(\delta,T)^{-1}$ for~$\delta =
2.66/10^4 \text{ eV}$.}
\end{figure}

The~graphs of~$k_B T$ and~$t(\delta,T)^{-1}$ for~$\delta = 2.66 \times
10^{-4}\text{ eV}$ are~plotted in~Fig.~\ref{fig:3}. Close similarity of~the~two plots
above~$15\text{ K}$ shows that~Eq.~\eqref{rho_T_ham} provides an~extension
of~\mbox{Kondo's}~formula~\eqref{Delta_rho} to~the~vicinity of~$0 \text{ K}$.

\section{\label{sec:3}Mean-field impurity heat~capacity}
Expressions for~the~mean-field energy~$\Delta U_{\text{s-d}}$
and~heat~capacity~$\Delta C$ of~a~dilute \mbox{s-d}~system, relative
to~that~of~a~free electron~gas, will~be~derived, using
the~Hamiltonian~$h^{(n,M)}(\xi,\eta)$, and~compared with~experimental data
on~impurity heat~capacity of~CuFe\cite{franck,mattis},
$\text{(LaCe)Al}_2$\cite{bader} and~CuCr\cite{triplett}. The~system's inverse
temperature~$\beta$ will~be~replaced by~$t(\delta,T)$. To~compensate deficiencies
of~$H_{\text{s-d}}$, such as~non-inclusion of~the~interaction between d-electrons
present in~the~\mbox{Anderson}~Hamiltonian, 
a~shift~$\Delta T > 0$ in~the~temperature scale 
will~be~implemented~(cf.~Refs.~\onlinecite{bloomfield,bader}). Thus~the~effective
inverse temperature~(EIT)
\begin{equation}
  t(\delta, T + \Delta T) = \delta^{-1} \tanh \left( \delta (k_B (T+
  \Delta T))^{-1}\right).
\end{equation}
\subsection{\label{subsec:3:1}CuFe alloys}
The~spin of~the~Fe~ions in~CuFe equals~$1/2$ according to~Ref.~\onlinecite{mattis}, 
therefore
\begin{equation}
  f_2(\xi) = \frac{\gamma}{M\sqrt{n_1}} \tanh \left( t(\delta, T+\Delta T)\gamma
  \sqrt{n_1} \xi \right)
\end{equation}
where~$n=Mn_1$, $\gamma = \sqrt{M} g$, $n_1$~denoting the~number of~conduction
electrons per~impurity. $f_1$~defined by~Eq.~\eqref{f1} is~a~linear function
in~the~simplest approximation:~$f_1(\xi) = b_0 + b_1 \xi + \ldots$\cite{m3}. 

The~expectation energy of~a~spin~$1/2$ system~$S_{\text{imp}}$ containing~$M$
impurities with~the~1-impurity Hamiltonian~\eqref{h_imp} equals
\begin{equation}
  U_{\text{imp}} = \left\langle h_{\text{imp}}^{(M)} \right\rangle_{h_{\text{imp}}
  ^{(M)}} = - M^2 n_1 \xi f_2(\xi) + \frac{1}{2}\gamma^2
  \label{U_imp}
\end{equation}
and, according~to~Appendix~B of~Ref.~\onlinecite{m3}, the~interaction energy
of~electrons with~the~Hamiltonian~$\tilde{h}_{\text{e}}^{(n,M)}(\xi,\eta)$ equals
\begin{equation}
  \Delta U_{\text{e}} = \left\langle \tilde{h}_{\text{e}}^{(n,M)}(\xi,\eta) 
  \right\rangle_{\tilde{h}_{\text{e}}^{(n,M)}(\xi,\eta)} - \left\langle n \Gamma_1^n
  H_0^{(1)} \right\rangle_{n \Gamma_1^n H_0^{(1)}} = - M n f_2(\xi) (b_0 + b_1
  f_2(\xi) + \ldots ),
  \label{DeltaU}
\end{equation}
where~$\xi$ is~the~minimizing solution of~Eq.~\eqref{xi}.
  
Using~Eqs.~\eqref{h_e},~\eqref{U_imp},~\eqref{DeltaU} and~the~definition
\begin{equation}
  h^{(n,M)}(\xi,\eta) := h_{\text{imp}}^{(M)}(\xi) + h_{\text{e}}^{(n,M)}(\xi,\eta)
\end{equation}
one~obtains
\begin{equation}
  \Delta U_{\text{s-d}} = U_{\text{imp}} + \Delta U_{\text{e}} + M n \xi f_2(\xi) - 
  \frac{1}{2} M n f_2^2(\xi).
  \label{DeltaU_sd}
\end{equation}
The~$n$-electron, $M$-impurity spin~$1/2$ \mbox{s-d}~system will~be now treated
as~a~subsystem of~a~sample~$S$ containing one mole of~impurities.
The~energy~$\Delta U_{\text{S}} = 6.022\times 10^{23} M^{-1}\Delta U_{\text{s-d}}$
of~such sample expressed in~mcals, equals
\begin{equation}
  \Delta U_{\text{S}} = 602.2\times 38271.78 M^{-1} \left( \frac{1}{2}\gamma^2 -
  \frac{1}{2} M^2 n_1 f_2^2(\xi) - M^2 n_1 f_2(\xi) (b_0 + b_1 f_2(\xi) + \ldots)
  \right),
\end{equation}
if~$b_0$, $\gamma$, $\delta$, $\xi$~are~given in~powers of~eV.

\begin{figure}
  \includegraphics{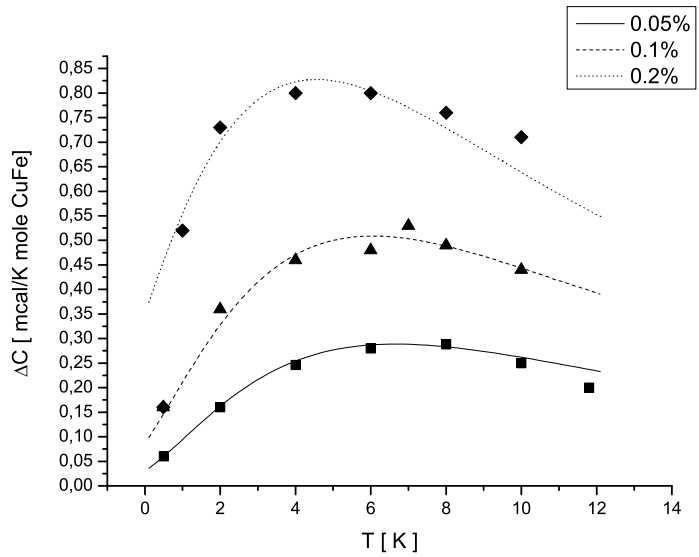}
  \caption{\label{fig:4}Impurity heat capacity~$\Delta C$ of~CuFe given
  by~Eq.~\eqref{DeltaC}, with~$b_0$, $b_1$, $\gamma$, $M$, $\delta$, $\Delta T$~equal
  to~the~values in~Table~\ref{tab:1}. The~points are~experimental results
  from~Refs.~\onlinecite{franck,mattis}.}
\end{figure}

The~excess heat~capacity of~one~mole of~CuFe alloy, relative to~that~of~one~mole 
of~pure~Cu, then~equals
\begin{equation}
  \Delta C = n_1^{-1} \left( \frac{\partial \Delta U_{\text{S}}}{\partial T} + 
  \frac{\partial \Delta U_{\text{S}}}{\partial \xi} \frac{\partial \xi}{\partial T} 
  \right), 
  \label{DeltaC}
\end{equation}
where~$\xi$~is~the~unique minimizing solution of~Eq.~\eqref{xi} and
\begin{equation}
  \frac{\partial \xi}{\partial T} = - \frac{\partial f_3}{\partial T}\left( \frac{
  \partial f_3}{\partial \xi} - 1 \right)^{-1}.
  \label{partial_xi}
\end{equation}
The~mean-field~$\Delta C(T + \Delta T)$ curves best fitting to~experimental data
of~Refs.~\onlinecite{franck,mattis} were~obtained for~the~values of~$M$, $\gamma$,
$b_0$, $b_1$, $\delta$, $\Delta T$~given in~Table~\ref{tab:1} and~are~depicted
in~Fig.~\ref{fig:4}. Agreement with~experiment is~good, especially for~$c=0.05 \%$,
$0.1\%$ and below~$10\text{ K}$. Discrepancies at~higher temperatures are~presumably
due~to~experimental error, which increases
with~temperature~(e.g.~Refs.~\onlinecite{bader,triplett}), and~to~an~increase
of~the~spin values of~Fe~ions in~this~temperature range\cite{bloomfield,daybell}.
In~fact, \mbox{Triplett}~et~al.\cite{triplett} estimate the~spin of~Fe~ions in~CuFe
to~be~equal~$3/2$.

\begin{table}
  \caption{\label{tab:1}}
  \begin{ruledtabular}
    \begin{tabular}{c|c|c|c|c|c|c|c|c}
      Alloy & $c$ & $n_1$ & $b_0[10^{-3}\sqrt{\text{eV}}]$ & $b_1$ & $\gamma
      [\sqrt{\text{eV}}]$ & $M$ & $\delta [10^{-4}\text{ eV}]$ & $\Delta T [\text{K}]$ \\
      \hline
      CuFe & $0.05 \%$ & $2000$ & $0.1$ & $-101$ & $0.19$ & $10^8$ & $7.14$ &
      $2.6$ \\
      \hline
      CuFe & $0.1 \%$ & $1000$ & $0.1$ & $-101$ & $0.245$ & $10^9$ & $7.48$ &
      $3$ \\
      \hline
      CuFe & $0.2 \%$ & $500$ & $0.1$ & $-101$ & $0.288$ & $10^{10}$ & $7.48$ &
      $3.9$ \\
      \hline
      CuCr & $212\times 10^{-7}$ & $10^7/212$ & $1.01/n_1$ & $-631$ & $0.086$ & 
      $36000$ & $10^{-13}$ & $0.78$ \\
      \hline
      CuCr & $51\times 10^{-6}$ & $10^6/51$ & $1.09/n_1$ & $-461$ & $0.091$ & 
      $248500$ & $10^{-13}$ & $1.05$ \\
      \hline
    \end{tabular}
    \end{ruledtabular}
\end{table}

\begin{figure}
  \includegraphics{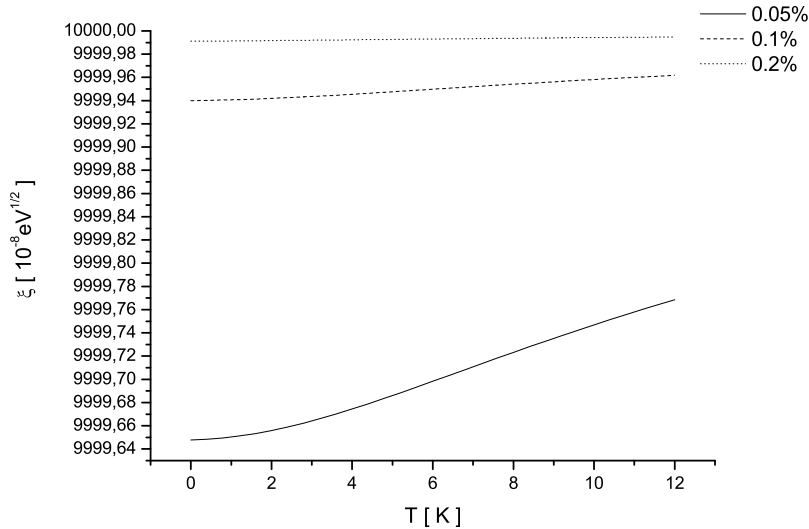}
  \caption{\label{fig:5}The~plots of~the~minimizing solutions~$\xi(T+\Delta T)$
  of~Eq.~\eqref{xi} corresponding to~the~$\Delta C(T+\Delta T)$ curves of~CuFe
  depicted in~Fig.~\ref{fig:4}.}
\end{figure}

The~minimizing solutions of~Eq.~\eqref{xi} are~plotted in~Fig.~\ref{fig:5}.
Uniqueness of~these~solutions is~proved in~the~Appendix.

It~is~worth emphasizing how~significantly variation of~$c$ affects the~shape of~both
experimental and~theoretical plots of~$\Delta C(T + \Delta T)$ in~Fig.~\ref{fig:4}.
It~has~been~suggested by~some~authors~(e.g.~Refs.~\onlinecite{triplett,brock})
that~nonlinearity of~$\Delta C$ in~$c$ observed in~DMA is~due to~impurity-impurity
interactions. The~above analysis shows that~this~nonlinearity can~be~explained
on~the~grounds of~the~\mbox{s-d}~Hamiltonian without additional interaction terms.

Another remarkable property of~the~theoretical plots of~$\Delta C(T + \Delta T)$
in~Fig.~\ref{fig:4} is~their~dependence on~$M$. The~best fitting 
values~$M_{\text{f}}$ of~$M$ fall in~the~range~$1 \ll M_{\text{f}} \ll A = 6.022 
\times 10^{23}$. The~sample~$S$ can~be therefore viewed as~made~up of~magnetic 
domains, each~containing~$M_{\text{f}}$ impurities with~a~favoured impurity-spin 
orientation, which~varies, in~general, from~one domain to~another. Experiment 
has~confirmed existence of~magnetic domains in~some~magnetic 
materials~(e.g.~Ref.~\onlinecite{aharoni}). The~orientation of~electron spins 
is~opposite to~that~of~impurities, as~follows 
from~Eqs.~\eqref{h_e_tilde},~\eqref{h_imp}.

Since the~solution~$\xi(T)$ of~Eq.~\eqref{xi} decreases
with~decreasing~$T$~(Fig.~\ref{fig:5}), it~follows that~the~ordering of~impurity
spins declines as~the~temperature is~lowered. A~similar dependence of~$\xi(T)$
on~$c$ can~be~observed. In~order to~find~$\lim \xi(T)$ as~$T\to 0$ for~small
enough~$c$ let~us recall that~$\lim t(\delta, T)=\beta$ as~$\delta \to 0$. The~graphs
of~$\xi(T)$ for~$t=\beta$, depicted in~Fig.~3 of~Ref.~\onlinecite{m3}, as~well as~the
form of~Eq.~\eqref{xi} in~all cases considered therein, show that as~$T\to 0 $,
$\lim \xi(T) = 0$ if~$\delta = 0$, $\Delta T = 0$. Hence, there~is no~ordering
of~impurity spins at~$T=0$ if~$\delta = 0$, $\Delta T = 0$, a~picture which agrees
with~the~RKKY~description of~interactions between impurity spins in~a~DMA.
\subsection{\label{subsec:3:2}$\text{(LaCe)Al}_2$}
\mbox{Bader}~et~al.\cite{bader} performed interesting measurements of~$\Delta C / c$
on~$(\text{La}_{1-x}\text{Ce}_x)\text{Al}_2$, with~$x=0.0064$, in~external magnetic 
fields ranging from~$0 \text{ kOe}$ to~$38 \text{ kOe}$. For strong fields 
\mbox{Andrei}~et.~al., obtained good agreement of~their~1-impurity~$\Delta C(T)$ 
function with~rescaled data of~Ref.~\onlinecite{bader}. 

According to~Refs.~\onlinecite{felsch,bader}, a~typical \mbox{Kondo}~effect, without
any~superconducting side-effects observed in~$(\text{La}_{1-x}\text{Ce}_x)\text{Al}_2$ 
with~Ce content exceeding~$x=0.0067$. However, for~$x=0.0064$, \mbox{Bader}~et.~al. 
estimate the~difference between the~expected normal~state and~measured 
superconducting-state heat~capacities as~insignificant. Thus~a~mean-field normal-state 
theory of~$\Delta C/c$ for~$(\text{La}_{1-x}\text{Ce}_x)\text{Al}_2$, with~$x=0.0064$, 
can~be~expected to~provide good agreement with~experiment.

The~number of~valence electrons per~host atom in~LaAl will~be~assumed~$8/3$. 
For~$x=0.0064$, 
\begin{equation*}
  c=\frac{0.0064}{2.9936} = \frac{4}{1871}, \qquad n_1 = \frac{8}{3} c^{-1} =
  \frac{3742}{3}.
\end{equation*}
The~ground~state of~the~Ce ion can~be~described by~a~spin with~magnitude~$1/2$
and~a~modified~\mbox{Lande}~factor~$g' = 10/7$~(Ref.~\onlinecite{felsch}), therefore
in~the~presence of~an~external magnetic field~$H$,
the~system's~Hamiltonian is
\begin{equation}
  H_{\text{s-d}}^{(n,M)}(H) = H_{\text{s-d}}^{(n,M)} - \frac{1}{2} g' \mu_B
  \tilde{H} \sum_{\alpha} S_{z\alpha} - \frac{1}{2} \mu_B H A^{(n)} \sum_i
  \mathbb{I}_0 \otimes \sigma_{zi},
  \label{H_sd_H}
\end{equation}
$\mathbb{I}_0$~denoting the~identity in~$L^2(\mathbb{R}^3)^n$ and
\begin{equation}
  f_2(\xi) = \frac{\gamma}{M\sqrt{n_1}} \tanh \left( t(\delta, T + \Delta T)
  (\gamma \sqrt{n_1}\xi - \frac{1}{2}g' \mu_B \tilde{H})\right),
  \label{f_2_H}
\end{equation}
where~$\tilde{H}=g_0 H$ is~the~effective magnetic field at~each~impurity site
and~$\mu_B$ the~\mbox{Bohr}~magneton. The~excess energy~$\Delta U_S/c$
of~a~sample~$S$ of~$\text{(LaCe)Al}_2$, with~$x=0.0064$, expressed in~joules, equals
\begin{equation}
  \begin{split}
  c^{-1} \Delta U_S & = \frac{1}{4} M^{-1} 1871 \times 602.2 \times 160.2 \times \\
  & \left( \frac{1}{2} \gamma^2 - \frac{1}{2} M^2 n_1 f_2(\xi)^2 - M^2 n_1 f_2(\xi) 
  \left( \xi - f_2(\xi) \right) + \frac{1}{2} \gamma^{-1} \mu_B g' \tilde{H} M^2 
  \sqrt{n_1} f_2(\xi) \right) \\
\end{split}
\label{Delta_USC}
\end{equation}
if~$b_0$, $\gamma$, $\delta$, $\xi$~are~expressed in~powers of~eV. For~one mole
of~impurities
\begin{equation}
  \Delta C = \frac{\partial \Delta U_S}{\partial T} + \frac{\partial \Delta
  U_S}{\partial \xi} \frac{\partial \xi}{\partial T}.
  \label{DeltaC_mole}
\end{equation}

\begin{figure}
  \includegraphics{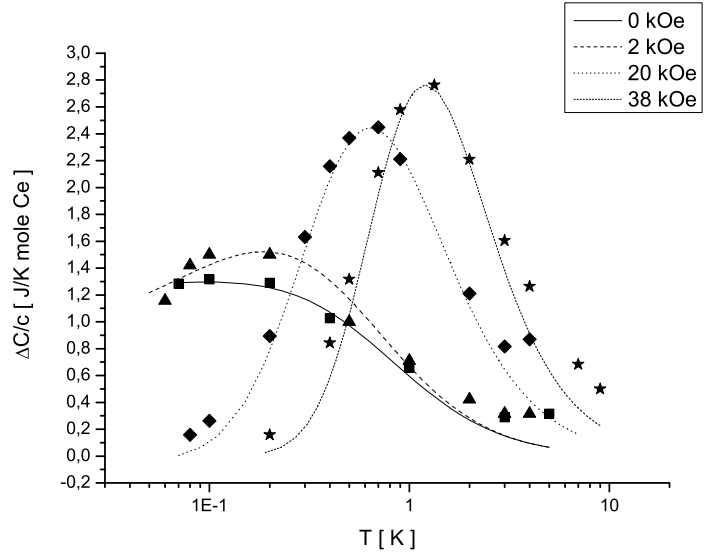}
\caption{\label{fig:6}The~plots of~$\Delta C(T+\Delta T)/c$
of~$(\text{La}_{1-x}\text{Ce}_x)\text{Al}_2$ in~various external magnetic fields,
as~given by~Eqs.~\eqref{DeltaC},~\eqref{Delta_USC}, for~$x=0.0064$ and~$b_0$,
$b_1$, $\gamma$, $M$, $\delta$, $\Delta T$, $g_0$~equal to~the~values
in~Table~\ref{tab:2}. The~points are~experimental results
from~Ref.~\onlinecite{bader}.}
\end{figure}

Adjusting the~parameters~$b_0$, $b_1$, $\gamma$, $M$, 
$\delta$, $\Delta T$, $g_0$, one~obtains the~best fitting to~experiment~$\Delta
C(T + \Delta T)/c$ curves plotted in~Fig.~\ref{fig:6}. The~corresponding values
of~the~parameters are~given in~Table~\ref{tab:2}. The~mean-field thermodynamics
founded on~the~Hamiltonian~$h^{(n,M)}$ thus~provides satisfactory agreement
with~experimental data on~the~field dependence of~$\Delta C(T)$.

\begin{table}
  \caption{\label{tab:2}}
  \begin{ruledtabular}
    \begin{tabular}{c|c|c|c|c|c|c|c|c}
      $b_0[10^{-4}\sqrt{\text{eV}}]$ & $b_1$ & $\gamma[10^{-4}\sqrt{\text{eV}}]$ & $M$ & 
      $\delta[10^{-5}\text{ eV}]$ & $\Delta T [\text{K}]$ & $g_0$ & $H [\text{kOe}]$ &
      $g^2[10^{-9}\text{ eV}]$\\
      \hline
      $1.9$ & $-13101$ & $3.1$ & $65$ & $3.1215$ & $0.39$ & - & $0$ & $1.4785$ \\
      \hline
      $1.9$ & $-13101$ & $3.1$ & $64$ & $0.6243$ & $0.27$ & $0.006$ & $2$ & $1.50156$ \\
      \hline
      $1.9$ & $-13101$ & $7.285$ & $215$ & $3.1215$ & $0.11$ & $0.006$ & $20$ &
      $2.46843$ \\
      \hline
      $1.9$ & $-13101$ & $10.85$ & $850$ & $12.486$ & $0.1$ & $0.006$ & $38$ &
      $1.38497$ \\
      \hline
    \end{tabular}
    \end{ruledtabular}
\end{table}

The~value of~$g_0$ was~found by~adjusting~$\Delta C(T+\Delta T)/c$ to~experiment
for~$H=2\text{ kOe}$, with~$b_0$, $b_1$, $\gamma$~equal to~their~best fitting values
for~$H=0$ and~allowing only small variations of~$M$. The~smallness of~$g_0$ indicates 
the~strong influence of~the~\mbox{Kondo}~effect in~the~formation of~a~polarization 
cloud of~conduction electrons around each~impurity. The~cloud screens each~magnetic ion 
from~interactions with other~magnetic ions\cite{mattis} and, as~implied 
by~the~inequality~$g_0 \ll 1$, also~from~the~applied field~$H$.
\subsection{\label{subsec:3:3}CuCr alloys}
\mbox{Triplett}~et.~al.\cite{triplett} performed high-precision measurements of~CuCr
impurity heat~capacity~$\Delta C$ for~a~variety of~concentrations.
For~$c=51\text{ ppm}$ their~$\Delta C(T)$ peak is~well defined and~terminates at~low
temperatures with~a~$\Delta C$ jump which they consider to~be~the~effect
of~impurity-impurity interactions. Explanation of~these~experimental~$\Delta C(T)$
data above the~jump temperature is~ a~challenge to~any~\mbox{s-d}~theory.

According to~\mbox{Monod}~et.~al.\cite{monod} the~spin of~the~Cr~ions in~CuCr
equals~$3/2$. Thus~one~finds
\begin{equation}
  f_2(\xi) = \frac{\gamma}{M\sqrt{n_1}} \frac{3 \mathrm{e}^{-4t\gamma^2 M^{-1}}\sinh
  \left( 3t\gamma \sqrt{n_1}\xi \right) + \sinh \left( t\gamma \sqrt{n_1}\xi
  \right)}{\mathrm{e}^{-4t\gamma^2 M^{-1}}\cosh \left( 3t\gamma \sqrt{n_1}\xi \right) 
  + \cosh \left( t\gamma \sqrt{n_1}\xi \right)}
\end{equation}
and
\begin{equation}
  U_{\text{imp}} = \left\langle h_{\text{imp}}^{(M)} \right\rangle_{h_{
  \text{imp}}^{(M)}} = - M^2 n_1 \xi f_2(\xi) + \frac{1}{2} \gamma^2 + f_4(\xi)
  \label{Uimp_sd}
\end{equation}
where
\begin{equation*}
  f_4(\xi) = 4 \gamma^2 \frac{\mathrm{e}^{-4t\gamma^2 M^{-1}}\cosh \left( 3t\gamma 
  \sqrt{n_1}\xi \right)}{\mathrm{e}^{-4t\gamma^2 M^{-1}}\cosh \left( 3t\gamma 
  \sqrt{n_1}\xi \right) + \cosh \left( t\gamma \sqrt{n_1}\xi \right)}.
\end{equation*}
Using~Eqs.~\eqref{DeltaU},~\eqref{DeltaU_sd}, one~obtains the~following formula
for~$\Delta U_S$ (expressed in~joules) of~a~sample~$S$ of CuCr:
\begin{equation}
  \Delta U_S = 602.2 \times 160.2 \times M^{-1} \left( \frac{1}{2}\gamma^2 +
  f_4(\xi) - \frac{1}{2}M^2 n_1 f_2^2(\xi) - M^2 n_1 f_2(\xi)(\xi - f_2(\xi))
  \right),
  \label{Delta_US}
\end{equation}
if~$b_0$, $\gamma$, $\delta$, $\xi$~are given in~powers of~eV. $\Delta C(T+\Delta
T)$~obtains using~Eqs.~\eqref{partial_xi}, \eqref{DeltaC_mole}.

\begin{figure}
  \includegraphics{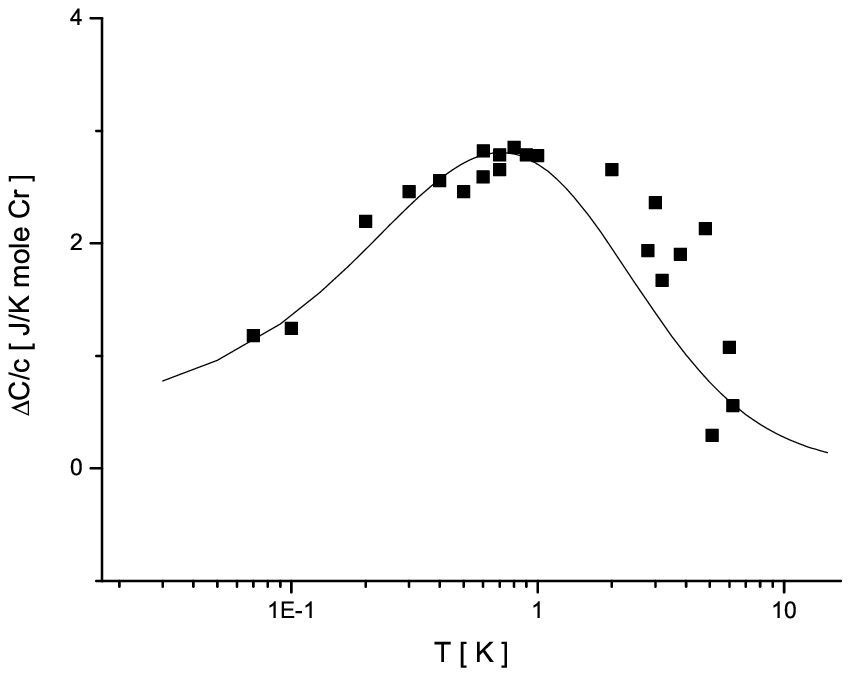}
  \caption{\label{fig:7}The~graph of~$\Delta C(T+\Delta T)/c$ of~CuCr
  with~$c=21.7\text{ ppm}$ as~given by~Eqs.~\eqref{DeltaC},~\eqref{Delta_US}, 
  with~$b_0$, $b_1$, $\gamma$, $M$, $\delta$, $\Delta T$~equal to~the~values
  in~Table~\ref{tab:1}. The~points are~experimental results
  from~Ref.~\onlinecite{triplett}.}
\end{figure}

\begin{figure}
  \includegraphics{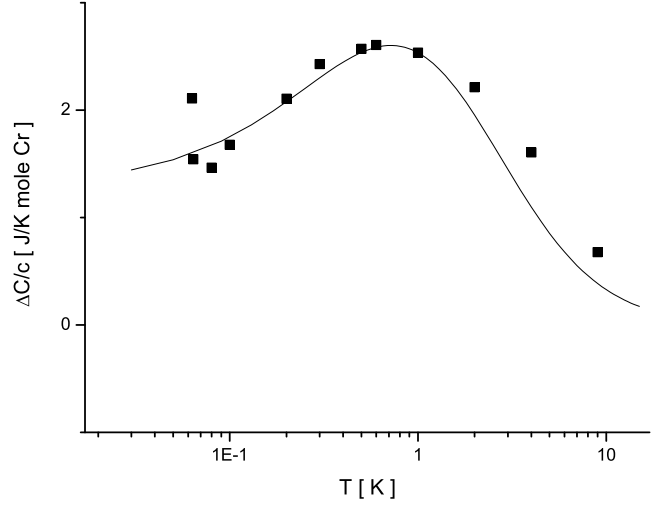}
  \caption{\label{fig:8}The~graph of~$\Delta C(T+\Delta T)/c$ of~CuCr
  with~$c=51\text{ ppm}$ as~given by~Eqs.~\eqref{DeltaC},~\eqref{Delta_US}, 
  with~$b_0$, $b_1$, $\gamma$, $M$, $\delta$, $\Delta T$~equal to~the~values
  in~Table~\ref{tab:1}. The~points are~experimental results
  from~Ref.~\onlinecite{triplett}.}
\end{figure}

The~best fitting graphs of~$\Delta C(T+\Delta T)/c$ for~$c=21.7\text{ ppm}$ and~$c=
51\text{ ppm}$ are~plotted in~Figs.~\ref{fig:7},~\ref{fig:8}. The~corresponding
values of~parameters~$b_0$, $b_1$, $\gamma$, $M$, $\delta$, $\Delta T$~are given
in~Table~\ref{tab:1}.

Agreement with~experimental data of~Ref.~\onlinecite{triplett} is~satisfactory,
especially for~~$c=21.7\text{ ppm}$, although not~as~good
as~for~CuFe~(Section~\ref{subsec:3:1}), presumably due~to~simplicity of~the~assumed
\mbox{s-d}~interaction in~Eq.~\eqref{H_sd} and~variation of~Cr~spin values at~higher
temperatures. It~has~been~suggested\cite{bloomfield,schrieffer} that~for~larger 
impurity spins the~\mbox{s-d}~interaction should~account~for the~momentum dependence
of~\mbox{s-d}~coupling.
\section{\label{sec:4}Magnetization of~$\text{(LaCe)Al}_2$}
Various measurements of~DMA impurity magnetization~$\Delta
M$~(e.g.~Ref.~\onlinecite{felsch} and~references therein) point to~a~similar field
dependence of~the~$\Delta M(H,T)$ vs.~$H/T$ curves. A~typical experimental plot 
of~$\Delta M(H,T)$, for~$(\text{La}_{1-x}\text{Ce}_x)\text{Al}_2$ with~$x=0.015$, 
can~be~found in~Ref.~\onlinecite{felsch}. The~single-impurity 
theoretical~$\Delta M(H,T)$ curves found by~\mbox{Andrei}~et~al.\cite{rajan,andrei}
fit the~rescaled data of~Ref.~\onlinecite{felsch} up~to~a~small error.

\begin{figure}
  \includegraphics{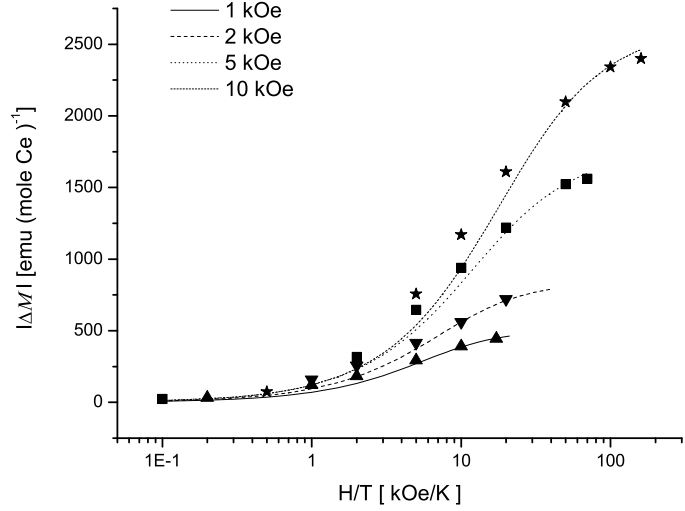}
  \caption{\label{fig:9}The~impurity magnetization~$|\Delta M|$ of~$(\text{La}_{1-x}
  \text{Ce}_x)\text{Al}_2$ in~various external magnetic fields as~given
  by~Eq.~\eqref{DeltaM}, with~$x=0.015$ and~$b_0$, $b_1$, $\gamma$, $M$, $\delta$,
  $\Delta T$, $g_0$~equal to~the~values in~Table~\ref{tab:3}. The~points
  are~experimental results from~Ref.~\onlinecite{felsch}.}
\end{figure}

As~implied by~the~form of~the~mean-field counterpart
of~the~Hamiltonian~\eqref{H_sd_H}, viz.,
\begin{equation*}
  h^{(n,M)}(H) = h^{(n,M)}(\xi,\eta) - \frac{1}{2} g' \mu_B \tilde{H} \sum_{\alpha} 
  S_{z\alpha} - \frac{1}{2} \mu_B H \sum_i \mathbb{I}_0 \otimes \sigma_{zi} 
  A^{(n)},
\end{equation*}
for~a~mole of~spin~$1/2$ impurities
\begin{equation}
  \Delta M = \frac{1}{2} g' \mu_B \sum_{\alpha=1}^A \left\langle S_{z\alpha}
  \right\rangle_{h_{\text{imp}}^{(A)}} = \frac{1}{2} g' \mu_B A \tanh\left(
  t(\delta, T+\Delta T) (\frac{1}{2} g' \mu_B \tilde{H} - \gamma \sqrt{n_1}\xi)
  \right)
  \label{DeltaM}
\end{equation}
where~$\xi$ is~the~unique solution of~Eq.~\eqref{xi} with~$f_2(\xi)$ given
by~Eq.~\eqref{f_2_H}. The~resulting plots of~$\Delta M(H,T)$ for~various applied
fields are~depicted in~Fig.~\ref{fig:9}. The~corresponding values of~$b_0$,
$b_1$, $\gamma$, $M$, $\delta$, $g_0$~are~presented in~Table~\ref{tab:3}. Agreement
with~experiment is~good in~the~range of~low temperatures, but~less satisfactory
at~higher~$T$. 

\begin{table}
  \caption{\label{tab:3}}
  \begin{ruledtabular}
    \begin{tabular}{c|c|c|c|c|c|c|c|c}
      $b_0[10^{-11}\sqrt{\text{erg}}]$ & $b_1$ & $\gamma[10^{-10}\sqrt{\text{erg}}]$ & 
      $M$ & $\delta[10^{-17}\text{erg}]$ & $\Delta T [\text{K}]$ & $g_0$ & 
      $H [\text{kOe}]$ & $g^2[10^{-24}\text{erg}]$\\
      \hline
      $28.8576$ & $-13101$ & $3.9237$ & $10^4$ & $2$ & $0.045$ & $0.006$ & $1$ &
      $15.3954$ \\
      \hline
      $38.4768$ & $-13101$ & $7.8474$ & $2\times 10^4$ & $3.1$ & $0.094$ & $0.006$ & $2$ 
      & $30.7908$ \\
      \hline
      $48.096$ & $-13101$ & $19.6185$ & $2.5\times 10^4$ & $4$ & $0.18$ & $0.006$ &
      $5$ & $153.9542$ \\
      \hline
      $48.096$ & $-13101$ & $39.237$ & $3\times 10^4$ & $4.2$ & $0.23$ & $0.006$ &
      $10$ & $513.1807$ \\
      \hline
    \end{tabular}
    \end{ruledtabular}
\end{table}

\section{\label{sec:5}Magnetic susceptibility}
The~zero-field impurity susceptibility
\begin{equation}
  \Delta \chi = \left( \frac{\partial \Delta M}{\partial \tilde{H}} + \frac{\partial
  \Delta M}{\partial \xi} \frac{\partial \xi}{\partial \tilde{H}}
  \right)_{\tilde{H}=0}
  \label{Delta_chi}
\end{equation}
has~been the~most frequently measured property of~DMA. The~theory of~$\Delta \chi$,
developed by~\mbox{Souletie}~et~al.\cite{souletie} for~a~DMA with~RKKY~interaction
between impurities, predicts a~dependence of~the~form
\begin{equation}
  \Delta \chi(T,c) = f(T/c)
  \label{Delta_chi_Tc}
\end{equation}
where~$f$ is~a~function independent of~concentration.
\mbox{Felsch}~et~al.\cite{felsch} have~confirmed approximate validity
of~Eq.~\eqref{Delta_chi_Tc} for~$(\text{La}_{1-x}\text{Ce}_x)\text{Al}_2$ with~$x$ 
ranging from~$0.02$ to~$0.06$.

Here validity of~formula~\eqref{Delta_chi}, with~$\Delta M$ given
by~Eq.~\eqref{DeltaM}, is~tested on~$\Delta \chi$ experimental data for~CuFe
with~$c=110\text{ ppm}$~(Ref.~\onlinecite{daybell}) and~$(\text{La}_{1-x}\text{Ce}_x)
\text{Al}_2$ with~$x=0.015$, $0.02$~(Ref.~\onlinecite{felsch}).
\subsection{\label{subsec:5:1}CuFe}
\begin{figure}
 \includegraphics{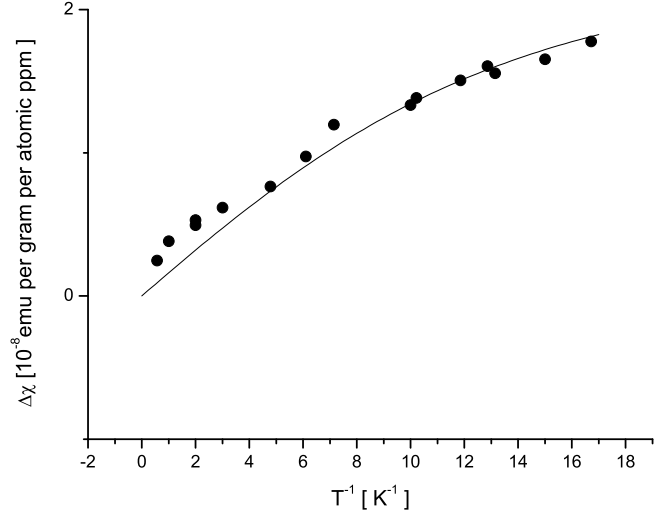}
\caption{\label{fig:10}The~impurity susceptibility~$\Delta \chi(T^{-1})$ of~CuFe,
with~$c=110 \text{ ppm}$, as~given by~Eq.~\eqref{Delta_chi}, with~$b_0$, $b_1$, 
$\gamma$, $M$, $\delta$, $g'$, $\Delta T$~equal to~the~values in~Table~\ref{tab:4}. 
The~points are~experimental results from~Ref.~\onlinecite{daybell}.}
\end{figure}

\mbox{Daybell}~et~al.\cite{daybell} expressed their~measured~$\Delta \chi$ values
for~CuFe in~emu~per~gram of~alloy per~ppm. Formula~\eqref{DeltaM}, expressed
in~these~units, takes the~form
\begin{equation}
  \Delta M_{\text{CuFe}} = \frac{1}{220} g' \mu_B M_{\text{Fe}} \tanh\left(
  t(\delta, T+\Delta T)(\frac{1}{2} g' \mu_B \tilde{H} - \gamma
  \sqrt{n_1}\xi)\right)
\end{equation}
where~$M_{\text{Fe}} = 1.042032405\times 10^{18}$ is~the~number of~Fe~ions contained in~one~gram
of~CuFe with~$c=110\text{ ppm}$. A~possible fit of~the~resulting function~$\Delta
\chi$, expressed in~these units, to~the~experimental data on~$\Delta \chi$, for~CuFe
from~Ref.~\onlinecite{daybell} is~presented in~Fig.~\ref{fig:10}. The~corresponding
values of~the~parameters are~given in~Table~\ref{tab:4}. Concavity of~the~$\Delta
\chi(T^{-1})$ curve in~Fig.~\ref{fig:10} appears not~to~be~fully adjustable
to~concavity of~the~experimental plot~at higher temperatures, however, agreement
is~satisfactory in~the~range of~low temperatures. The~discrepancy between theory 
and~experiment at~higher~$T$ is~presumably due~to~the~increase of~Fe~spin values
with~increasing~$T$~(Ref.~\onlinecite{daybell}).

\begin{table}
  \caption{\label{tab:4}}
  \begin{ruledtabular}
    \begin{tabular}{c|c|c|c|c|c|c|c|c|c|}
      Alloy & $x$ & $n_1$ & $b_0[10^{-10}\sqrt{\text{erg}}]$ & $b_1$ & $\gamma
      [10^{-15}\sqrt{\text{erg}}]$ & $M$ & $\delta [10^{-17}\text{ erg}]$ &
      $g'$ & $\Delta T [\text{K}]$ \\
      \hline
      CuFe & - & $10^5/11$ & $8$ & $-101$ & $1$ & $10^4$ & $1$ & $1.05$ &
      $0.005$ \\
      \hline
      $(\text{La}_{1-x}\text{Ce}_x)\text{Al}_2$ & $0.015$ & $1592/3$ & $2.4048$ & 
      $-13101$ & $392370$ & $200$ & $1$ & $10/7$ & $0.2$\\
      \hline
      $(\text{La}_{1-x}\text{Ce}_x)\text{Al}_2$ & $0.02$ & $1192/3$ & $2.4048$ & 
      $-13101$ & $392370$ & $400$ & $0.1$ & $10/7$ & $0.6$ \\
      \hline
    \end{tabular}
    \end{ruledtabular}
\end{table}

\begin{figure}
  \includegraphics{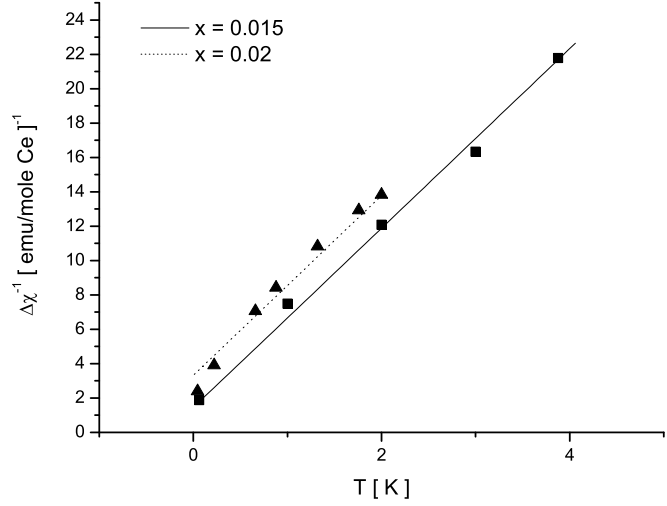}
  \caption{\label{fig:11}The~inverse impurity susceptibility~$\Delta \chi(T+ \Delta T
  )^{-1}$ of~$(\text{La}_{1-x}\text{Ce}_x)\text{Al}_2$ according
  to~Eq.~\eqref{Delta_chi}, with~$x$, $b_0$, $b_1$, $\gamma$, $M$, $\delta$, $g'$, 
  $\Delta T$~equal to~the~values in~Table~\ref{tab:4}. The~points are~experimental 
  results from~Ref.~\onlinecite{felsch}.}
\end{figure}

\subsection{\label{subsec:5:2}$\text{(LaCe)Al}_2$}
\mbox{Felsch}~et~al.\cite{felsch} performed detailed measurements 
of~$(\text{La}_{1-x}\text{Ce}_x)\text{Al}_2$ susceptibility for~$x$ ranging from~$0.01$
to~$0.2$. Their~plots of~$\Delta \chi(T)^{-1}$ for~$x=0.01$, $0.015$ are almost
indistinguishable and~follow a~\mbox{Curie}-\mbox{Weiss}~law
\begin{equation}
  \Delta \chi (T) = \chi_{CF}(T) T / (T+\Theta), \qquad \chi_{CF} T = \text{const}
  \label{Delta_chi_T}
\end{equation}
for~$T \in (0.15\text{ K}, 3\text{ K})$. Below~$0.15\text{ K}$ their data on~$\Delta 
\chi(T)^{-1}$ for~$x=0.015$, deviate from~Eq.~\eqref{Delta_chi_T} to~lower values,
contrary to~measurements of~$\Delta \chi$ on~AuV\cite{dam} and earlier
theories\cite{anderson1,anderson2,schotte,wilson} which~predict a~flattening-off 
of~$\Delta \chi(T)^{-1}$ to higher values. For~$x \ge 0.02$ the~$\Delta 
\chi(T)^{-1}$ curves of~Ref.~\onlinecite{felsch} no~longer 
obey~Eq.~\eqref{Delta_chi_T} and~exhibit a~weak concavity.

Eq.~\eqref{Delta_chi}, with~$\Delta M$ expressed by~Eq.~\eqref{DeltaM}, provides
a~good fit to~the~$\Delta \chi(T+\Delta T)^{-1}$ data of~Ref.~\onlinecite{felsch}
for~$x=0.01$, $0.015$ and, as~could~be~expected from~Eq.~\eqref{free_en}, a~less
satisfactory adjustment for~$x \ge 0.02$. The~plots of~$\Delta \chi(T+\Delta T)^{-1}$,
as~given by~Eq.~\eqref{Delta_chi}, are~depicted in~Fig.~\ref{fig:11} for~$x=0.015$,
$0.02$. A~minor deviation from~the~\mbox{Curie}-\mbox{Weiss}~law at~very low
temperatures, similar to~the~one~found for~AuV in~Ref.~\onlinecite{dam},
can~be~observed. The~corresponding values of~$b_0$, $b_1$, $\gamma$, $M$,
$\delta$~are given in~Table~\ref{tab:4}.
\section{\label{sec:6}Concluding remarks}
The~mean-field theory of~dilute \mbox{s-d}~systems presented
in~Ref.~\onlinecite{m2} has, in~general, proved successful in~providing quantitative
explanation of~the~$T$, $c$, $H$~dependence of~DMA heat~capacity, magnetization
and~susceptibility in~the~range of~low temperatures.

It~has~been~shown that~nonlinear dependence of~DMA impurity heat~capacity
on~$c$ can~be~explained in~the~dilute limit exclusively in~terms
of~the~\mbox{s-d}~interaction, without introducing an~impurity-impurity potential. 

The~EIT~$t$ has~improved the~dependence of~all~thermodynamic functions on~temperature
and~removed the~singularity in~\mbox{Kondo's}~expression for~DMA impurity
resistivity.

Deviations of~the~theory from~experiment, in~the~case of~heat~capacity, 
magnetization and~CuFe susceptibility, are~presumably due to~increasing spin values of~impurity ions
observed in~some DMA at~higher temperatures and simplicity of~the~\mbox{s-d}~coupling
assumed. It~has~been~suggested that, apart~from~\mbox{s-wave}, the~coupling
should~account~for the~\mbox{d-type} character of~the~interaction.

Further improvement of~the~theory can~be~expected after including higher expansion
terms  of~$f_1$ and~correction terms to~the~mean-field free~energy. The~variation
of~impurity spin values at~higher temperatures also~suggests investigation
of~the~thermodynamics of~a~more general \mbox{s-d}~system containing impurity spins
with various~$S=1/2,1,3/2,\ldots$.
\appendix*
\section{}
To~prove uniqueness of~the~minimizing solutions of~the~equations
\begin{equation}
  \frac{\partial}{\partial \xi} f(h^{(n,M)}(\xi,\eta),\beta) = 0, 
  \label{po_xi}
\end{equation}
\begin{equation}
  \frac{\partial}{\partial \eta} f(h^{(n,M)}(\xi,\eta),\beta) = 0, 
  \label{po_eta}
\end{equation}
it~suffices to~note that~Eqs.~\eqref{po_xi}, \eqref{po_eta} take
the~form~(cf.~Ref.~\onlinecite{m2})
\begin{equation}
  \xi = f_1(\xi - \eta) + f_2(\xi),
\end{equation}
\begin{equation}
  \eta = f_1(\xi - \eta).
\end{equation}
Thus, $\eta(\xi) = \xi - f_2(\xi)$, and~therefore, for~$\eta$ satisfying these
equations\cite{m2},
\begin{equation}
  \frac{\mathrm{d}}{\mathrm{d}\xi} f(h^{(n,M)}(\xi,\eta(\xi)),\beta) =
  \frac{\partial}{\partial \xi} f(h^{(n,M)}(\xi,\eta(\xi)),\beta) = M (\xi -
  f_1(f_2(\xi)) - f_2(\xi)).
\end{equation}
Since~$f_1(x) = b_0 + b_1 x$ and~$b_1 < 0 $, $1 \ll |b_1|$, $f_2'(\xi) > 0$,
it~follows that the~solution~$\xi_m$ of~Eq.~\eqref{xi} is~unique and
\begin{equation*}
   \frac{\mathrm{d}}{\mathrm{d}\xi} f(h^{(n,M)}(\xi,\eta(\xi)),\beta) < 0 \qquad
   \text{for } \xi < \xi_m,
\end{equation*}
\begin{equation*}
   \frac{\mathrm{d}}{\mathrm{d}\xi} f(h^{(n,M)}(\xi,\eta(\xi)),\beta) > 0 \qquad
   \text{for } \xi > \xi_m.
\end{equation*}
\bibliography{Mackowiak}
\end{document}